\documentclass[runningheads]{llncs}
\usepackage[utf8]{inputenc}
\usepackage{amsmath,amssymb,amsfonts}
\usepackage{listings,enumitem,tikz,multirow,tcolorbox}
\usepackage{centernot,multicol,float,url,graphics}
\usepackage[final]{pdfpages}
\usepackage[colorlinks=true,urlcolor=blue]{hyperref}
\usepackage[framemethod=tikz]{mdframed}

\usepackage[framemethod=tikz]{mdframed}

\usepackage{array}
\usepackage{booktabs}
\usepackage{graphicx}
\usepackage{threeparttable}
\usepackage{wasysym} 
\usepackage{multirow}

\usepackage[framemethod=tikz]{mdframed}

\makeatletter
\def\ProblemSpecBox{
  \@ifnextchar[\ProblemSpecBox@opt{\ProblemSpecBox@noopt}}
\def\ProblemSpecBox@opt[#1]#2{
  \protected@edef\@currentlabelname{#1}
  \protected@edef\@currentlabel{#1}
  \begin{mdframed}[
    innerlinewidth=0.5pt,
    innerleftmargin=10pt,
    innerrightmargin=10pt,
    innertopmargin = 10pt,
    innerbottommargin=10pt,
    skipabove=\dimexpr\topsep+\ht\strutbox\relax,
    roundcorner=5pt,
    frametitle={#2},
    frametitlerule=true,
    frametitlerulewidth=1pt]
}
\def\ProblemSpecBox@noopt#1{
  \ProblemSpecBox@opt[#1]{#1}
}
\def\endProblemSpecBox{
  \end{mdframed}
}
\makeatother

\begin{document}


\title{\textsc{\mdseries Flatee}: Federated Learning Across Trusted Execution Environments\thanks{This paper appeared at IEEE EuroS\&P 2021 as Poster~\cite{mondal2021poster}.}}
\titlerunning{FLATEE}

\author{Arup Mondal \and Yash More \and Ruthu Hulikal Rooparaghunath \and Debayan Gupta}
\authorrunning{A. Mondal et al.}
%
\institute{Ashoka University, Sonipat, Haryana, India
\email{\newline arup.mondal\_phd19@ashoka.edu.in, yashrajmore29@gmail.com, hulikalruthu@gmail.com, debayan.gupta@ashoka.edu.in}}

\maketitle


\begin{abstract}
Federated learning allows us to distributively train a machine learning model where multiple parties share local model parameters without sharing private data. However, parameter exchange may still leak information.  Several approaches have been proposed to overcome this, based on multi-party computation, fully homomorphic encryption, etc.; many of these protocols are slow and impractical for real-world use as they involve a large number of cryptographic operations. In this paper, we propose the use of Trusted Execution Environments (TEE), which provide a platform for isolated execution of code and handling of data, for this purpose. We describe \textsc{Flatee}, an \textit{efficient} privacy-preserving federated learning framework across TEEs, which considerably reduces training and communication time. Our framework can handle malicious parties (we do not natively solve adversarial data poisoning, though we describe a preliminary approach to handle this).

\end{abstract}

\section{Introduction}\label{section:introduction}
While traditional machine learning approaches depend on a central training data set, privacy considerations have driven interest in decentralized learning frameworks, where parties collaborate to train an ML model without sharing their respective training datasets. Federated learning (FL)~\cite{mcmahan2016communication} is a powerful approach for collaborative and privacy-preserving learning: here, parties collectively train a model by training locally and then exchanging model parameters (instead of actual training data), which keeps their data private.
However, recent work~\cite{melis2019exploiting} has demonstrated that parameter interaction and the final model may leak information about the training dataset.


Multi-party computation has been used for privacy-preserving ML~\cite{ramachandran2021s++}. Made possible by a range of cryptographic primitives, MPC~\cite{mood2016frigate,perry2014systematizing,di2014practical} allows multiple parties to compute a function without revealing the inputs of any individual party (beyond what is implied by the output). Several schemes have already been proposed for privacy-preserving FL using MPC~\cite{truex2019hybrid}, but these often take a very long time to train, and may also incur high data-transmission costs. Further, they cannot usually deal with participants dropping out during the FL process~\cite{truex2019hybrid}. 

Trusted Execution Environments (TEE) are an emerging hardware primitive: Intel’s Software Guard Extensions (SGX)~\cite{gupta2016using,DBLP:journals/iacr/CostanD16} provide a module within chipsets that enable the creation of secure containers called \textit{enclaves}. These hardware-enforced ``reverse sandboxes'' allow data and code to be processed without the influence of code running in the traditional registers of the processor. An SGX system can use hardware-based attestation to prove that an enclave executes exactly the functions promised and nothing else (assuming one trusts Intel). TEEs incur lower overheads compared to traditional software protections.

\subsection{Contributions}

We propose an efficient, privacy-preserving federated learning framework using a TEE (Intel SGX). We have an FL server $\mathcal{S}$ and a set of parties $\mathcal{P}=P_1, P_2, \dots, P_n$ where each $P_i$ has a private dataset $\mathbb{D}_i$.
Critically, unlike recent MPC or Fully Homomorphic Encryption (FHE) based solutions, we do \textbf{not} exchange data -- however obfuscated -- with the server. Instead, we use traditional FL techniques modified for privacy, training models separately within each party/client TEE, and then combining them securely within the server. Specifically:
\begin{itemize}
    \item \textsc{Flatee}, a privacy-preserving federated learning framework based on TEEs,  enables the parties to efficiently train a distributed model. Additionally, \textsc{Flatee} provides strict privacy guarantees of training   and is also resistant to \textit{data-poisoning} and \textit{model-poisoning} attacks.

    \item We use Differential Privacy (DP) based techniques with low clipping bounds and high noise variance to prevent backdoor attacks. We also use Multi-KRUM~\cite{blanchard2017machine} to guarantee resiliency from $k$ malicious updates out of $n$ total updates, where $2k+2 <n$.

        

    
\end{itemize}
This paper is organized as follows: Sec~\ref{section:preliminaries} provides background on FL and TEEs. Sec~\ref{section:flatee} describes our framework and its trust model. Sec~\ref{section:futurework} discusses future work.


\subsection{Motivation}

We address two main drawbacks of existing frameworks. Popular FL frameworks such as~\cite{bonawitz2017practical,truex2019hybrid} often incur large amounts of training time and communication latencies (due to the computationally intensive cryptographic operations involved). Privacy-preserving federated learning frameworks are vulnerable to post-quantum attacks, as they only involve traditional encryption techniques. Our proposed framework relies on quantum-resilient cryptographic schemes which prevent such potential attacks and protects user-sensitive data. Furthermore, our TEE-based approach offers a significant time improvement over others, as shown in Table~\ref{table:CompAnalysis}.    

\begin{table*}[ht]
\caption{Comparison between various privacy-preserving federated learning framework}
\label{table:CompAnalysis}
\centering
\scriptsize
\newcommand*\rot[1]{\hbox to1.1em{\hss\rotatebox[origin=br]{-60}{#1}}}
\newcommand*\feature[1]{\ifcase#1 $\square$\or$\boxtimes$\or$\blacksquare$\or\Circle\or\LEFTcircle\or\CIRCLE\or$\checkmark$\or$\times$\or \fi}
\newcommand*\e[3]{\feature#1&\feature#2&\feature#3}
\newcommand*\g[2]{\feature#1&\feature#2}
\newcommand*\f[5]{\feature#1&\feature#2&\feature#3&\feature#4&\feature#5}
\newcommand*\h[4]{\feature#1&\feature#2&\feature#3&\feature#4}
\makeatletter
\newcommand*\ex[8]{#1\tnote{#2}&#3&\g#4&\e#5&\g#6&\g#7&\f#8&\expandafter\h\@firstofone
}
\makeatother
\newcolumntype{E}{c@{}c@{}c}
\newcolumntype{G}{c@{}c}
\newcolumntype{F}{c@{}c@{}c@{}c@{}c}
\newcolumntype{H}{c@{}c@{}c@{}c}
\begin{threeparttable}
\begin{tabular}{@{}lc G  E  G  G F H@{}}
\toprule
Framework  & Comms$^{\dag}$ & \multicolumn{2}{c}{\shortstack{Privacy \\ Capability}} & \multicolumn{3}{c}{\shortstack{Threat \\ Model}} & \multicolumn{2}{c}{\shortstack{Privacy \\ Guarantees}} & \multicolumn{2}{c}{\shortstack{Security \\ Guarantees}} & \multicolumn{5}{c}{\shortstack{Techniques \\ Used}} & \multicolumn{4}{c}{Features}\\
\midrule

&& \rot{Inference}
 & \rot{Training}
 & \rot{Participants}
 & \rot{Aggregator}
 & \rot{Required TPA}
 & \rot{Computation}
 & \rot{Output}
 & \rot{Data Poisoning}
 & \rot{Model Poisoning}
 & \rot{HE}
 & \rot{TP}
 & \rot{SS+AE}
 & \rot{FE}
 & \rot{TEE}
 & \rot{Scalability}
 & \rot{Dynamic Participants}
 & \rot{Post-Quantum Security} 
 & \rot{Statistical Heterogeneity} \\
\midrule

\ex{PySyft~\cite{ryffel2018generic}}{} {2mn+n} {35} {016} {55} {33} {53333} {4333} \\

\midrule

\ex{Truex et al.~\cite{truex2019hybrid}}{} {2mt+mn+n} {35} {216} {55} {33} {35333} {5333} \\

\midrule

\ex{Bonawitz et al.~\cite{bonawitz2017practical}}{} {2mn+n} {35} {216} {55} {33} {33533} {5433} \\

\midrule

\ex{HybridAlpha~\cite{xu2019hybridalpha}}{} {mn+m+n} {45} {216} {55} {43} {33353} {5533} \\

\midrule

\ex{\textsc{\mdseries Flatee}}{} {mn} {55} {217} {55} {55} {33335} {5555} \\

\bottomrule
\end{tabular}

\begin{tablenotes}
\item ``HE`` is homomorphic encryption'; ``TP'' is Threshold-Paillier system; ``SS+AE'' secret sharing with key agreement protocol and authenticated encryption scheme; ``FE'' is functional encryption; and ``TEE'' is Trusted Execution Environment. ``TPA'' is \textit{trusted third party} is used to set up a master private key and a master public key that will be used to derive multiple public keys to one or more parties who intend to encrypt their data. \feature0  denotes honest party; \feature1 denotes semi-honest party; \feature2 denotes dishonest party;  \feature3 denotes does not provides property; \feature4 denotes partially provides property;  \feature5 denotes provides property. Communications\textsuperscript{\dag} -- The number of crypto-related operations required in each training round, where $n$ is the number of participants and $m$ is the number of aggregators, and $t$ is the threshold for decryption of Threshold-Paillier cryptosystem.
\end{tablenotes}      

\end{threeparttable}
\end{table*}


\section{Technical Background and Preliminaries}\label{section:preliminaries}

\subsection{Federated Learning}\label{subsection:fl}
Federated learning (FL)~\cite{mcmahan2016communication} is a distributed approach to Machine Learning which allows models to be trained on a large body of decentralized data with many participants. FL is an example of the technique of ``bringing code to data, not data to code'', and is suitable for use cases with sensitive data (health care, financial services, etc.). 
In FL, each party trains a model locally and exchanges only model parameters with an FL \textit{server} or \textit{aggregator}, instead of the private training data.

The participants in the training processes are \textit{parties} and the \textit{FL server}, which is a cloud-based distributed service. Devices agreement to the server that they are ready to run an \textit{FL task} for a given \textit{FL population}. An FL population is specified by a globally unique name which identifies the learning problem, or application, which is worked upon. An FL task is a specific computation for an FL population, such as training to be performed with given hyperparameters, or evaluation of trained models on local device data. After finishing the local computation on its local dataset then each device updates the model parameters (e.g. the weights of a neural network) to the FL server. The server incorporates these updates into its global state of the global model.

\subsection{Trusted Execution Environment}\label{subsection:tee}
A Trusted execution environment (TEE) is a hardware extension that aims to provide \textit{integrity} and \textit{confidentiality} guarantees to security-sensitive computation performed on a computer where all the privileged software (kernel, hypervisor, etc) is potentially malicious. Specifically~\cite{gupta2016using,DBLP:journals/iacr/CostanD16}: 

\begin{enumerate}
    \item \textit{Authenticity \& Confidentiality} of the code running on a TEE is ensured.
    
    \item The \textit{State Integrity} of run-time states is also ensured including memory, CPU registers, and I/O; states are stored in persistent memory.
    
    \item The content of a TEE is \textit{dynamic} and can be updated during execution.
    
    \item An ``ideal'' TEE is \textit{secure} against all software and hardware attacks.
    
    \item A TEE is \textit{trustworthy} and can provide proof of correctness of the executed computation to a third-party.
    
    \item Provides proof that users are interacting with software hosted inside the TEE (the \textit{attestation} functionality).
\end{enumerate}

A major aim of TEEs is to solve the problem of secure remote computation -- execution on an untrusted machine while having integrity and trust guarantees. One example of such a system is Intel SGX~\cite{gupta2016using,DBLP:journals/iacr/CostanD16}, which provides a secure container using trusted hardware to give a remote user the ability to upload the code and data to this container. Several measures ensure the confidentiality of the executed computation and intermediate data. 

\begin{definition}[Secure Enclave] 
A TEE consists of a \textit{Processor Reserved Memory} (PRM) system which contains \textit{Enclave Page Cache} (EPC) which has multiple designated memory pages to store data and code. Each such page refers to a distinct secure enclave. A secure enclave has two different attributes, associated data and associated code. We will represent an enclave as $\mathbb{E}(data, code)$ in the following definitions. \textbf{Enclave measurement}, $M(\mathbb{E}(data, code))$, is the hash of the data and code placed inside the enclave including ordering and positioning.
\end{definition}

\begin{definition}
A TEE can be described as these following algorithms.
\begin{itemize}
    \item $\mathsf{TEE.create()}\rightarrow \mathsf{\mathbb{E}(\emptyset, \emptyset)}$ creates a new, uninitialized enclave from a free EPC page.
    \item $\mathsf{TEE.add(\mathbb{E}(\emptyset, \emptyset), data, code)\rightarrow\mathbb{E}(data, code)}$ loads the data and the associated code to the enclave.
    \item $\mathsf{TEE.extend(\mathbb{E}(data, code))\rightarrow M(\mathbb{E}(data, code))}$ to update the measurements of the
    enclave. The enclave's measurements $\mathsf{(M(\mathbb{E(.,.)})}$ are used by a remote party for attestation purposes. Any connecting remote party would compare the expected measurement and trusted hardware reported measurement to establish trust.
    \item $\mathsf{TEE.init(\mathbb{E}(data, code))\rightarrow HMAC(M(\mathbb{E}(data, code)))}$ sets initialization of the enclave to true, opens way for the loaded code to be executed and finalizes the hash of the measurements of the enclave.
    \item $\mathsf{TEE.KeyDerive(\mathbb{E}(data, code))\rightarrow key, Enc_{pk}}$. Derives symmetric encryption key for transfer of enclave associated data and code and the public key associated with this particular enclave.
    \item $\mathsf{TEE.remove(\mathbb{E}(data, code))}\rightarrow \mathbb{E}(\emptyset, \emptyset)$ clears all the assigned memory and deassigns processing power assigned to the initialized enclave.
\end{itemize}
\end{definition}

\section{\textsc{\mdseries Flatee} Framework}\label{section:flatee}
\begin{figure}
    \centering
    \includegraphics[width=0.7\textwidth]{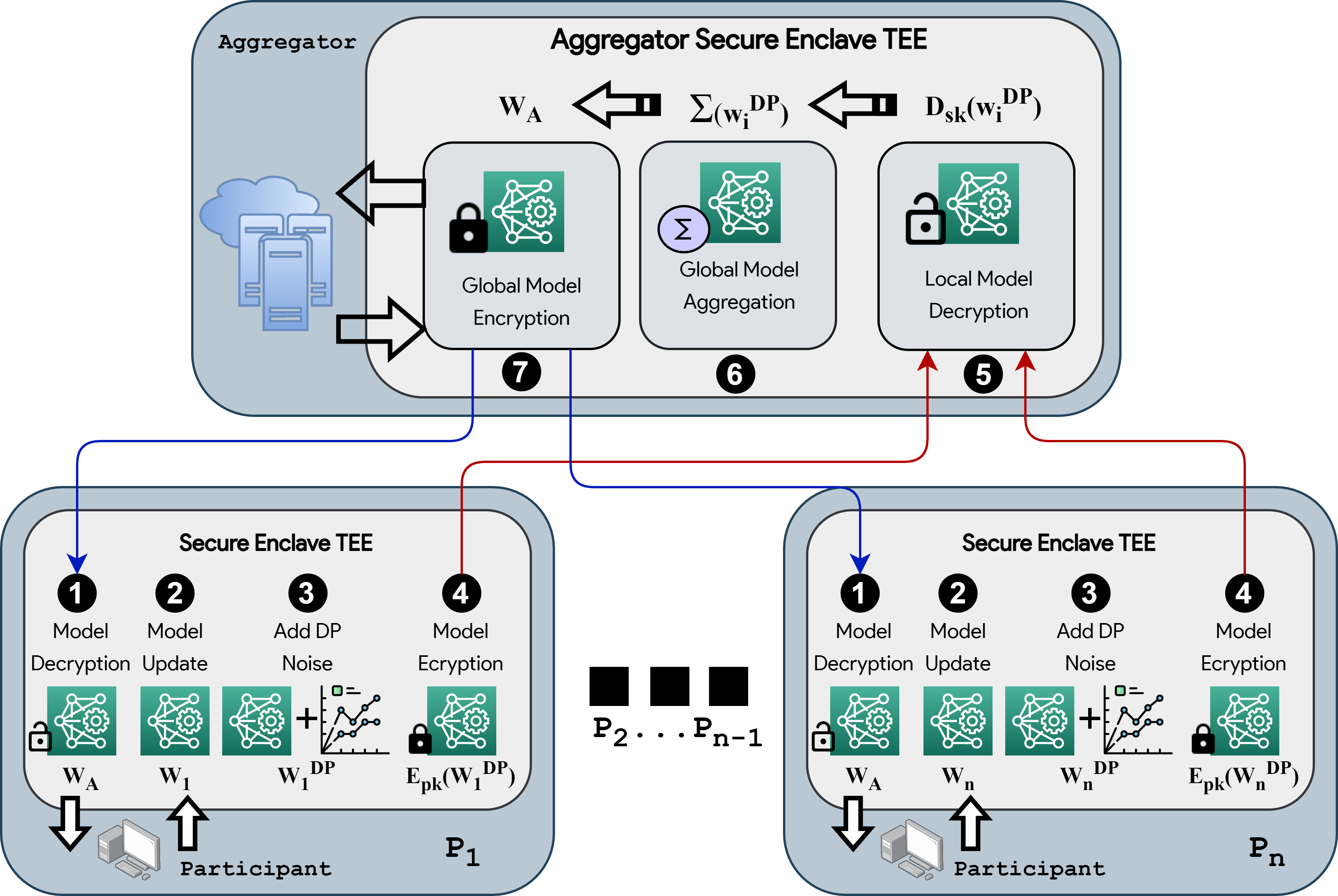}
    \caption{Schematic overview of \textsc{Flatee}}
    \label{figure:flatee}
\end{figure}

Let $\mathcal{S}$ be the FL server and $\mathcal{P}$ be a set of \textit{n} parties, where $i^{th}$ party $P_i$ holds its own private dataset $\mathbb{D}_i$, and $\mathcal{M}_{FL}$ is the DL model to be trained by the parties' private data. Each party agrees on $\mathcal{M}_{FL}$ before starting the training process and authenticates to the FL server. We assume an honest-but-curious, non-colluding FL server, which runs the protocol honestly but may try to glean information from the trained models. \textit{Curious, colluding participants} may inspect messages exchanged between the FL server or final model to glean the private data of other participants (we discuss the malicious case later).



\subsection{\textsc{\mdseries Flatee} Detailed Operations}\label{subsection:flateeop}

\subsubsection{Setup.}
$\mathcal{S}$ and all $\mathcal{P}$ have SGX-enabled machines, and agree to train a model $\mathcal{M}_{FL}$. $\mathcal{S}$ uses its SGX module to authenticate all $\mathcal{P}$ and aggregate the parties' trained models using the \textit{federated average function}~\cite{mcmahan2016communication} (\textit{e.g.,} weighted mean, geometric median etc.) to generate the global trained model. Each $P_i$ checks the model then trains it locally on private data, then sends the \textit{encrypted model parameters} to $\mathcal{S}$. $E$ and $D$ are \textit{post-quantum secure} en/decryption; ``$pk$'' and ``$sk$'' denote public and private keys (private keys never leave SGX).

\begin{mdframed}[font=\footnotesize,frametitle={\textsc{\mdseries Flatee} Protocol}]
\begin{enumerate}
    \item Each $P_i$ agrees on a model $\mathcal{M}_{FL}$ and $\mathcal{S}$ publishes a hash of the model $\mathcal H(M_{FL})$ by which everyone can verify their local version of the model.
    
    \item To ensure $P_i$ trains $\mathcal{M}_{FL}$, we check the sign measurement $TEE_{measure}^{P_i}$  signed by the the $P_i$ TEE's private key $TEE_{key_{P_i}}^{sk}$. $P_i$ also authenticates its TEE with  TEE of $\mathcal{S}$. 
    
    \item $P_i$ trains the local model and adds DP noise to the model parameters, and then sends the encrypted model parameters (using $TEE_{key_{FL server}}^{pk}$) to  $\mathcal{S}$. 
    
    \item Encrypted model parameters from all $P_i$, $\{E(\mathcal{M}_{FL}^{\mathbb{D}_{P_1}}), E(\mathcal{M}_{FL}^{\mathbb{D}_{P_2}}), \dots ,E(\mathcal{M}_{FL}^{\mathbb{D}_{P_n}})\}$ are decrypted inside of the FL server TEE's using $TEE_{key_{FL server}}^{sk}$, \textit{i.e.}, we perform $\{D(E(\mathcal{M}_{FL}^{\mathbb{D}_{P_1}})), D(E(\mathcal{M}_{FL}^{\mathbb{D}_{P_2}})), \dots ,D(E(\mathcal{M}_{FL}^{\mathbb{D}_{P_n}}))\}$.
    
    \item $\mathcal{S}$ runs the \textit{federated average function} in its TEE to aggregate all $\mathcal{P}$'s trained model parameters and get a global trained model. $\mathcal{S}$ should use \textit{data-
    obliviousness} (\textit{e.g.,} Oblivious RAM) to hide the actual memory reference sequence.
    
    \item $\mathcal{S}$ calculates the loss function over the global model. If it satisfies error constraints, $\mathcal{S}$ sends the encrypted global model $E(\mathcal{M}_{G})$ to $\mathcal{P}$, else we do another round -- steps 3 to 5. A party can drop out at any time of the training process but can join only after a training round. 
\end{enumerate}
\end{mdframed}


\subsubsection{Threat Model and Poisoning Attacks.}
We assume \textit{curious and colluding participants}. We separately consider adversarial participants who contribute poisoned updates to introduce backdoors into the shared model: here, assume that the adversary intends to harm the performance, or introduce backdoors, into the shared \textit{global} model, or leak private information about the used training dataset. For this work, we limit the adversaries to label flipping attacks, pixel-pattern backdoor attacks, and deep leakage from gradients using reconstruction attacks.

We use Multi-KRUM~\cite{blanchard2017machine}, a byzantine-resilient gradient aggregation algorithm, to address poisoning attacks. Instead of using a publicly available validation dataset, it scores each local model parameter based on its deviation from every other submitted local model parameters in every federated round. Multi-KRUM~\cite{blanchard2017machine} guarantees resiliency from $k$ malicious updates out of $n$ total updates, when $2k+2 <n$. For update $V_i \ \forall\  i\in [1, n]$, $Score(V_i)$ is calculated as the sum of euclidean distances between $V_i$ and $V_j$, where $V_j$ denotes the $n-k-2$ closest vectors to $V_i$ as follows: $Score(V_i) = \sum_{i\rightarrow j} \mid\mid V_i - V_j \mid\mid^2$. Here, $i \rightarrow j$ denotes the fact that $V_j$ belongs to $n-k-2$ closest vectors to $V_i$. The $n-k$ updates with the lowest scores are selected for aggregation, and the rest are discarded. Multi-KRUM offers ease of implementation and provides defense against model replacement attacks. In the future, we plan to compare aggregation techniques that defend against poisoning attacks, such as FoolsGold, median, and trimmed mean.
We refer the reader to~\cite{blanchard2017machine} for security guarantees and convergence analysis of Multi-KRUM. 

DP ~\cite{dwork2014algorithmic} based techniques with low clipping bounds and high noise variance can render backdoor attacks ineffective, but can have a slight impact on the accuracy of the \textit{global} model. Hence, to minimize the efficacy of model poisoning attacks with DP, but without impacting model performance, we can use gradient pruning where gradients with small magnitudes are pruned to zero.

\section{Future Work}\label{section:futurework}
In this work, we assume existence of ideal TEEs which are not affected by any micro-architectural attacks like \textit{Spectre}, \textit{Meltdown}, \textit{Foreshadow}, \textit{Plundervolt}, etc. We continue to work towards including cryptographic techniques to present a micro-architectural attack resistant protocol for the same. We also plan to present a complete implementation of our proposed framework whilst considering all known micro-architectural attack vectors.  
Finally, we also plan to compare the efficiency and cost of \textsc{Flatee} with existing protocols achieving similar results. 

\bibliographystyle{splncs04.bst}
\bibliography{arxiv-ref.bib}

\begin{thebibliography}{10}
\providecommand{\url}[1]{\texttt{#1}}
\providecommand{\urlprefix}{URL }
\providecommand{\doi}[1]{https://doi.org/#1}

\bibitem{blanchard2017machine}
Blanchard, P., El~Mhamdi, E.M., Guerraoui, R., Stainer, J.: Machine learning
  with adversaries: Byzantine tolerant gradient descent. In: Proceedings of the
  31st International Conference on Neural Information Processing Systems. pp.
  118--128 (2017)

\bibitem{bonawitz2017practical}
Bonawitz, K., Ivanov, V., Kreuter, B., Marcedone, A., McMahan, H.B., Patel, S.,
  Ramage, D., Segal, A., Seth, K.: Practical secure aggregation for
  privacy-preserving machine learning. In: Proceedings of the 2017 ACM SIGSAC
  Conference on Computer and Communications Security. pp. 1175--1191 (2017)

\bibitem{DBLP:journals/iacr/CostanD16}
Costan, V., Devadas, S.: Intel {SGX} explained. {IACR} Cryptology ePrint
  Archive  \textbf{2016}, ~86 (2016), \url{http://eprint.iacr.org/2016/086}

\bibitem{di2014practical}
Di~Crescenzo, G., Feigenbaum, J., Gupta, D., Panagos, E., Perry, J., Wright,
  R.N.: Practical and privacy-preserving policy compliance for outsourced data.
  In: International Conference on Financial Cryptography and Data Security. pp.
  181--194. Springer (2014)

\bibitem{dwork2014algorithmic}
Dwork, C., Roth, A., et~al.: The algorithmic foundations of differential
  privacy.  (2014)

\bibitem{gupta2016using}
Gupta, D., Mood, B., Feigenbaum, J., Butler, K., Traynor, P.: Using intel
  software guard extensions for efficient two-party secure function evaluation.
  In: International Conference on Financial Cryptography and Data Security. pp.
  302--318. Springer (2016)

\bibitem{mcmahan2016communication}
McMahan, H.B., Moore, E., Ramage, D., Hampson, S., et~al.:
  Communication-efficient learning of deep networks from decentralized data.
  arXiv preprint arXiv:1602.05629  (2016)

\bibitem{melis2019exploiting}
Melis, L., Song, C., De~Cristofaro, E., Shmatikov, V.: Exploiting unintended
  feature leakage in collaborative learning. In: 2019 IEEE Symposium on
  Security and Privacy (SP). pp. 691--706. IEEE (2019)

\bibitem{mondal2021poster}
Mondal, A., More, Y., Rooparaghunath, R.H., Gupta, D.: Poster: Flatee:
  Federated learning across trusted execution environments. In: 2021 IEEE
  European Symposium on Security and Privacy (EuroS\&P). pp. 707--709. IEEE
  Computer Society (2021)

\bibitem{mood2016frigate}
Mood, B., Gupta, D., Carter, H., Butler, K., Traynor, P.: Frigate: A validated,
  extensible, and efficient compiler and interpreter for secure computation.
  In: 2016 IEEE European Symposium on Security and Privacy (EuroS\&P). pp.
  112--127. IEEE (2016)

\bibitem{perry2014systematizing}
Perry, J., Gupta, D., Feigenbaum, J., Wright, R.N.: Systematizing secure
  computation for research and decision support. In: International Conference
  on Security and Cryptography for Networks. pp. 380--397. Springer (2014)

\bibitem{ramachandran2021s++}
Ramachandran, P., Agarwal, S., Mondal, A., Shah, A., Gupta, D.: S++: A fast and
  deployable secure-computation framework for privacy-preserving neural network
  training. arXiv preprint arXiv:2101.12078  (2021)

\bibitem{ryffel2018generic}
Ryffel, T., Trask, A., Dahl, M., Wagner, B., Mancuso, J., Rueckert, D.,
  Passerat-Palmbach, J.: A generic framework for privacy preserving deep
  learning. arXiv preprint arXiv:1811.04017  (2018)

\bibitem{truex2019hybrid}
Truex, S., Baracaldo, N., Anwar, A., Steinke, T., Ludwig, H., Zhang, R., Zhou,
  Y.: A hybrid approach to privacy-preserving federated learning. In:
  Proceedings of the 12th ACM Workshop on Artificial Intelligence and Security.
  pp. 1--11 (2019)

\bibitem{xu2019hybridalpha}
Xu, R., Baracaldo, N., Zhou, Y., Anwar, A., Ludwig, H.: Hybridalpha: An
  efficient approach for privacy-preserving federated learning. In: Proceedings
  of the 12th ACM Workshop on Artificial Intelligence and Security. pp. 13--23
  (2019)

\end{thebibliography}


\end{document}